\documentclass{article}
\usepackage{spconf,amsmath,graphicx}
\usepackage{multirow}
\usepackage{footnote} 
\usepackage{subfigure} 
\usepackage{url}
\usepackage{cite,balance}
\usepackage{graphicx, multirow, float}
\usepackage{xcolor}
\usepackage{multirow}
\newcommand{\fig}[1]{Fig.~}
\newcommand{\tab}[1]{Tab.~}
\newcommand{\Sec}[1]{Sec.~}
\newcommand{\eq}[1]{Eq.~}

\title{MFA: TDNN with Multi-scale Frequency-channel Attention \\ for Text-independent Speaker Verification with Short Utterances}
\name{Tianchi Liu$^{1,2}$, Rohan Kumar Das$^{3}$, Kong Aik Lee$^{1}$ and Haizhou Li$^{2,4}$\thanks{This work is partially supported by Science and Engineering Research Council, A$^\star$STAR, Singapore, through the National Robotics Program under Human-Robot Interaction Phase 1 (Grant No. 192 25 00054); in part by A$^\star$STAR under its Feasibility Study Scheme (Project No. FS-2021-001); in part by A$^\star$STAR under its RIE2020 Advanced Manufacturing and Engineering Domain (AME) Programmatic Grant No. A1687b0033.
}}

\address{$^{1}$Institute for Infocomm Research, A$\star$STAR, Singapore\\
$^2$Department of Electrical and Computer Engineering, National University of Singapore, Singapore\\
$^{3}$Fortemedia Singapore, Singapore~~~~~~
$^{4}$The Chinese University of Hong Kong, Shenzhen, China}
\begin{document}
\maketitle

\begin{abstract}

The time delay neural network (TDNN) represents one of the state-of-the-art of neural solutions to text-independent speaker verification. However, they require a large number of filters to capture the speaker characteristics at any local frequency region. In addition, the performance of such systems may degrade under short utterance scenarios. To address these issues, we propose a multi-scale frequency-channel attention (MFA), where we characterize speakers at different scales through a novel dual-path design which consists of a convolutional neural network and TDNN. We evaluate the proposed MFA on the VoxCeleb database and observe that the proposed framework with MFA can achieve state-of-the-art performance while reducing parameters and computation complexity. Further, the MFA mechanism is found to be effective for speaker verification with short test utterances.

\end{abstract}

\vspace{-1mm}
\vspace{-2mm}

\section{Introduction}
\vspace{-1mm}
Speaker verification (SV) aims to verify the claimed identity of an enrolled speaker, in either a text-dependent or a text-independent manner~\cite{campbell1997speaker}. The former is based on fixed short phrases~\cite{tdref}, while the latter doesn't impose any restrictions on the phonetic content~\cite{liu2022neural}. As the text is fixed in text-dependent SV, only a small amount of data is often used during training~\cite{das2018speaker}. On the contrary, text-independent SV requires a larger amount of training data than text-dependent SV~\cite{rkd_thesis}, that limits the scope of its applications. Data-efficient  text-independent SV for short utterances remains a challenge~\cite{liu2020text, das2018speaker, bhattacharya2017deep, huang2018angular}.

In the past few decades, SV research witnessed several major breakthroughs. One of the advanced frameworks for SV is x-vector~\cite{snyder2017deep, lee2021xi}, which employs time delay neural networks (TDNNs) to extract frame-level features and finally represent them by fixed-length vectors. 
Similarly, deep convolutional neural networks (CNN) were exploited and achieved highly competitive performance for SV~\cite{liu2019unified, yi2021perceptual}. Specially, ResNet-based networks have been widely adopted as the backbone for SV system in the recent years~\cite{ zhou2021resnext, li2017deep, desplanques2020ecapa, zhou2020dynamic, chung2020defence, tao2020audio, thienpondt2021integrating}.

Along the similar direction of neural speaker embedding, the ECAPA-TDNN~\cite{desplanques2020ecapa} achieves the current state-of-the-art (SOTA) performance by introducing more emphasis on channel attention, propagation and aggregation. However, TDNNs still suffer from the drawback whereby plenty of filters are required to model the speaker characteristics occurring at some local frequency regions~\cite{thienpondt2021integrating}. Meanwhile, their performance under short utterance scenario still needs attention.


We propose a multi-scale frequency-channel attention (MFA) module to handle the shortcomings of TDNNs to handle the short utterances for SV in this work. Motivated by ECAPA CNN-TDNN, which further strengthens ECAPA-TDNN by adding a CNN-based front-end to incorporate frequency translational invariance, the MFA module is designed as a front-end module for TDNNs, to obtain high resolution feature representations from short utterances. It employs a frequency-channel attention to assert the ability of effectively capturing speaker information from local regions for TDNNs. 

\begin{figure*}
\vspace{-3mm}
\centerline{\includegraphics[scale=0.09]{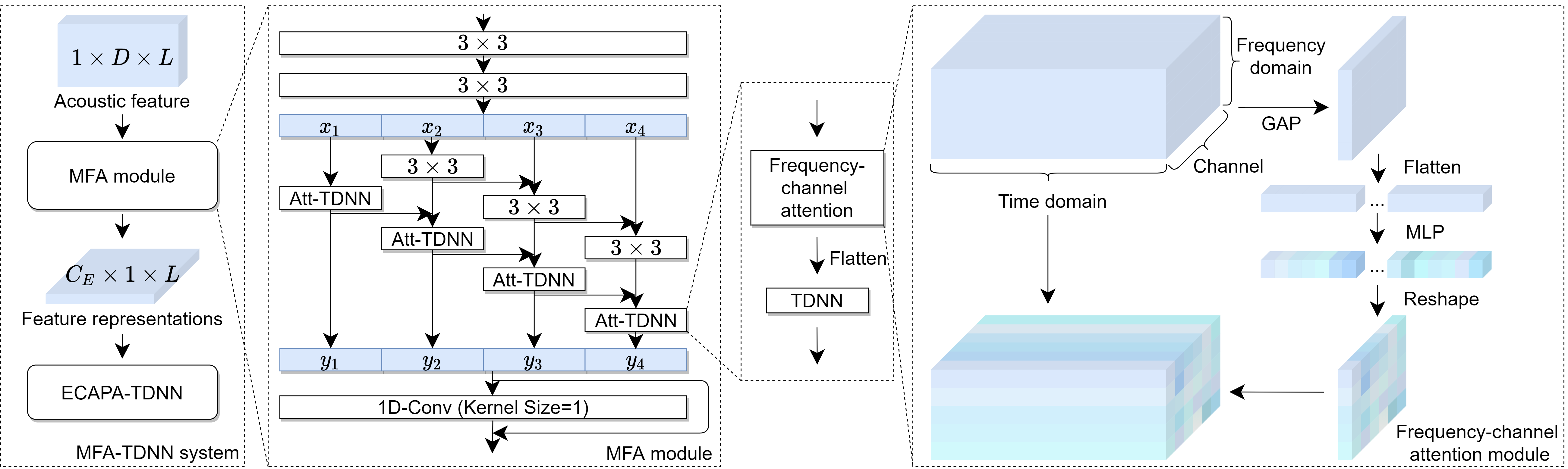}}
\vspace{-2mm}
\caption{A network architecture of the proposed MFA-TDNN system with multi-scale frequency-channel attention (MFA) module. The white box with 3 $\times$ 3 indicates 2D-CNN with kernel size of 3. The blue boxes indicate intermediate feature maps. We consider the scale dimension \textit{s}=4 as an example, therefore, the feature maps are equally divided into four groups along the channel dimension. The input acoustic feature is represented as a $1 \times D \times L$ feature map with 1 channel, $D$ dimensional feature frame and $L$ frames. $C_{E}$ indicates the number of channels in the convolutional frame layers of ECAPA-TDNN.
 } 
\label{fig:MFA}
\vspace{-3mm}
\end{figure*}

Further, the local and temporal information are modeled by a novel dual-path structure at multiple scales. Meanwhile, by incorporating these two modules, a local cross-channel attention mechanism is developed to contribute towards an efficient framework. The primary contributions of this work are: (1) proposal of frequency-channel attention module, (2) designing dual-pathway for local and temporal information modeling, and (3) local cross-channel attention for the MFA by incorporating (1) and (2), which altogether devote to have data-efficient SV framework for  short utterance scenarios.


\vspace{-2mm}
\section{Multi-scale Frequency-channel Attention}
\label{sec:MFA}
\vspace{-3mm}

The proposed MFA front-end is shown in \fig~\ref{fig:MFA}, which consists of two parts.  The dual-pathway multi-scale (DM) module extracts multi-scale information and the frequency-channel attention (FA) module is used to emphasize important local frequency region. By combining these two modules, we obtain an efficient local cross-channel attention mechanism.

\vspace{-2mm}
\subsection{Dual-pathway multi-scale module}
\label{sec:Dual-pathway}

The Res2Net has the ability to improve the representation by modelling multi-scale information within one layer~\cite{gao2019res2net}
. Recently, Res2Net is found to be very effective for SV task as well~\cite{zhou2021resnext, desplanques2020ecapa, thienpondt2021integrating}. In ~\cite{desplanques2020ecapa}, a Res2Net structure is studied where the 2D-CNN is replaced by a TDNN. In this way, the network is expected to process multi-scale features while re-scaling the frame-level features given a wider temporal context over the recording. 
However, as  short utterances offer limited information, it makes sense to extract the speaker characteristics at different scales for an improved speaker characterization. 


Motivated by the above, we propose a DM module for dealing with short utterance SV. Considering the fact that the TDNN-based Res2Net module relies on a large number of filters to process well across all the frequency bands and then models speaker characteristics occurring at any local frequency region, we enrich the representation ability of local regions by providing high-resolution details and emphasizing the important frequency bands. 

In this work, we divide the feature maps into several groups along the channel dimension \textit{C} based on the scale dimension factor \textit{s}. For the first pathway, a 2D-CNN with 3 $\times$ 3 filters of $n$ channels (\textit{n = C / s})  extracts features from group \textit{i} of input feature $x_{i}$. The output features of the previous 2D-CNN are then sent to next group \textit{i + 1} of 2D-CNN along with another group of input feature maps $x_{i+1}$. This pathway is expected to filter the high-resolution features from multi-scale regions. Similarly, the second pathway is designed between a small group of TDNNs to bridging the re-scaled frame-level features with global temporal information. In addition, the frequency channel attention is asserted between 2D-CNN and TDNN for each scale. It is expected to emphasize the specific frequency regions before feeding the feature into TDNN, which will be detailed in the next subsection.

Finally, a 1D-CNN is employed to fuse the information from different scales. The equivalent receptive field increases much more than the classic Res2Net due to the dual-pathway design for any possible path along with, where the input features are processed to output. This will result in many equivalent feature scales due to combination effects~\cite{gao2019res2net}.

\vspace{-3mm}
\subsection{Frequency-channel attention module}
With the advent of deep learning, various attention mechanisms are widely used in SV\cite{liu2020speaker, thienpondt2021integrating, desplanques2020ecapa, jiang2019effective}. Among them, squeeze-and-excitation (SE) block~\cite{hu2018squeeze} draws great attention due to its success in the field of computer vision, and has been applied to SV as well~\cite{thienpondt2021integrating, desplanques2020ecapa}. The information of some local frequency regions shows great importance for SV. Therefore, we inherit the idea of SE block and propose a novel FA block to emphasize the important information by re-weighting frequency bands and channels. 

The channel relationship represents a composition of instance-agnostic functions with local receptive regions~\cite{hu2018squeeze}. To effectively extract features from the local fields for short utterance SV, we attempt to re-scale local feature responses adaptively by modeling their interdependencies between both channels and frequency bands. Therefore, different from SE block, which only conducts attention on channels, we perform the attention across both channel and frequency dimensions. Specifically, the global temporal information is firstly squeezed into a channel-frequency descriptor as shown in \fig~\ref{fig:MFA}. This is achieved by using global average pooling (GAP). The adaptive re-scaling is performed by employing a multi-layer perceptron (MLP) following the excitation operation of SE block~\cite{hu2018squeeze}. Finally, the original feature map is re-weighted by the FA map obtained in the previous step.

Prior study shows that an efficient channel attention network (ECA-Net) is achieved by replacing the cross-channel interaction with the local cross-channel attention~\cite{wang2020eca}. By combining the FA module with a DM module in MFA, the channels are divided into multiple scales, thereby limiting the interactions only to adjacent channels. By doing so, we expect to improve the efficiency of the model in a similar way to ECA-Net, by avoiding the interaction across all channels.
\vspace{-2mm}
\section{Experiments}
\label{sec:EXPERIMENTS}
\vspace{-2mm}
\subsection{Database}
\label{sec:Database}
The experiments are conducted on VoxCeleb1~\cite{nagrani2017voxceleb} and VoxCeleb2 dataset~\cite{chung2018voxceleb2}. Only the development partition of the VoxCeleb2 dataset is used for training, and a small portion of about 2\% of it is reserved for validation. On the other hand, the VoxCeleb1 dataset is used for testing. It is noted that there are no overlapping speakers between the development set of VoxCeleb2 and VoxCeleb1 dataset. We evaluate the models on three different popular test set, namely VoxCeleb1-O, VoxCeleb1-E and VoxCeleb1-H. The full-length test utterances are used for when comparing with other SOTA systems.



As the training of neural networks benefits from data augmentation~\cite{park2019specaugment}, we employ five augmentation techniques to increase the diversity of the training data. The first two follow the idea of random frame dropout in the time domain~\cite{park2019specaugment} and speed perturbation~\cite{ko2015audio}. The remaining three are a set of reverberate data, noisy data, and a mixture of the both.


We build a truncated test set of short segments by keeping the original trial pairs, while considering only the first several seconds. We set the short duration cases in the range 4-10s, with a step of 1 s. As the shortest duration of the utterances in VoxCeleb1-O test set is 4s, those testing samples shorter than the set maximum duration are remaining unaltered.

\vspace{-2mm}
\subsection{Systems description}
We consider both ECAPA-TDNN~\cite{desplanques2020ecapa} and ECAPA CNN-TDNN~\cite{thienpondt2021integrating} as baseline systems for the studies. 
As we re-implement them on our own for the studies, the results of these two systems are marked with `(Re-implemented)' in Section~\ref{sec:results}. In our systems, we use 80-dimensional filterbank (Fbank) features. We provide details of the two baselines and proposed MFA-TDNN based systems in the following:
\begin{itemize}
\item 
{\bf ECAPA-TDNN (Re-implemented)}: It is the standard ECAPA-TDNN as proposed in~\cite{desplanques2020ecapa}. The number of channels is 512 in the convolutional frame layers.
\item 
{\bf ECAPA CNN-TDNN (Re-implemented)}: Four layers of CNN are employed as a front-end for ECAPA-TDNN as proposed in~\cite{thienpondt2021integrating}. Different from~\cite{thienpondt2021integrating}, we do not use a larger version of ECAPA-TDNN, but the standard version mentioned above. This is for fair comparisons with ECAPA-TDNN and proposed MFA-TDNN. 
\item
{\bf MFA-TDNN (Standard)}: The standard ECAPA-TDNN with proposed MFA as a front-end. The scale dimension \textit{s} is set to 4, while the dimension of channels \textit{C} for MFA is 32. 
\item
{\bf MFA-TDNN (Lite)}: The lite version of MFA-TDNN, where the number of channels is 480 instead of 512 in the convolutional frame layers of ECAPA-TDNN back-end. The scale dimension \textit{s} is set to 4, whereas the dimension of channels \textit{C} for MFA is 24.

\end{itemize}

\begin{table*}[]
\vspace{-3mm}
\caption{Performance in EER(\%) and minDCF of the baselines, proposed MFA-TDNN and other SOTA systems on VoxCeleb1 test sets. The MFCC indicates mel frequency cepstral coefficients and Spec. represents spectrogram.}
\label{tab:sota}
\vspace{1mm}
\setlength{\tabcolsep}{0.33mm}{
\begin{tabular}{cc|cc|cccccc}
\hline
\multirow{2}{*}{Model} & \multirow{2}{*}{Input Feature} & \multirow{2}{*}{\begin{tabular}[c]{@{}c@{}}Params\\ (Million)\end{tabular}} & \multirow{2}{*}{\begin{tabular}[c]{@{}c@{}}MAC \\ (Giga)\end{tabular}} & \multicolumn{2}{c}{VoxCeleb1-O} & \multicolumn{2}{c}{VoxCeleb1-E} & \multicolumn{2}{c}{VoxCeleb1-H} \\
 &  &  &  & EER & minDCF & EER & minDCF & EER & minDCF \\ \hline \hline
ECAPA-TDNN~\cite{desplanques2020ecapa} (Re-implemented) & 80-dim Fbank & 6.19 & 1.57 & 1.0050 & 0.0991 & 1.2076 & 0.1311 & 2.2593 & 0.2197\\
ECAPA CNN-TDNN\cite{thienpondt2021integrating} (Re-implemented) & 80-dim Fbank & 7.66 & 2.29 & 0.9199 & {0.0921} & 1.1107 & \textbf{0.1154} & 2.1406 & 0.2036\\ 
Proposed: \textbf{MFA-TDNN (Standard)} & 80-dim Fbank & \textbf{7.32} & \textbf{1.91} & \textbf{0.8561} & 0.0923 & \textbf{1.0834} & 0.1175 & \textbf{2.0487} & \textbf{0.1897}\\ 
Proposed: \textbf{MFA-TDNN (Lite)} & 80-dim Fbank & \textbf{5.93} & \textbf{1.50} & {0.9678} & \textbf{0.0913} & {1.1382} & {0.1208} & {2.1740} & {0.1993} \\ 
\hline\hline
Extended-TDNN (Large) ~\cite{desplanques2020ecapa} & 80-dim MFCC & 20.4 & - & 1.26 & 0.1399 & 1.37 & 0.1487 & 2.35 & 0.2153\\
Extended-TDNN~\cite{desplanques2020ecapa} & 80-dim MFCC & 6.8 & - & 1.49 & 0.1604 & 1.61 & 0.1712 & 2.69 & 0.2419\\
ResNet18~\cite{desplanques2020ecapa} & 80-dim MFCC & 13.8 & - & 1.47 & 0.1772 & 1.60 & 0.1789 & 2.88 & 0.2672\\
ResNet34~\cite{desplanques2020ecapa} & 80-dim MFCC & 23.9 & - & 1.19 & 0.1592 & 1.33 & 0.1560 & 2.46 & 0.2288\\
ARET-25~\cite{zhang2020aret} & 161-dim Spec. & 12.2 & - & 1.39 & - & 1.52 & - & 2.61 & -\\
H/ASP~\cite{kwon2021ins} & 64-dim Fbank & 8.0 & - & 1.15 & - & 1.35 & - & 2.49 & -\\
DDB + Gate~\cite{jiang2019effective} & 30-dim MFCC & 8.83 & - & 2.31 & 0.268 & - & - & - & -\\
\hline
\end{tabular}
\vspace{-3mm}
}
\end{table*}

\begin{table*}[]
\vspace{-2mm}
\caption{Performance in EER (\%) of baselines and proposed MFA-TDNN  under short utterance scenario on truncated testset.}
\label{tab:short}
\vspace{1mm}
\centering
\begin{tabular}{c|ccccccc}
\hline
\multirow{2}{*}{System} & \multicolumn{7}{c}{Duration for test utterances}                                                                           \\
                        & $\le$ 10 s       & $\le$ 9 s          &  $\le$ 8 s         &  $\le$ 7 s         &  $\le$ 6 s         &  $\le$ 5 s         &  $\le$ 4 s        \\ \hline\hline
ECAPA-TDNN              & 1.0050          & 0.9731          & 0.9784          & 0.9997          & 1.0529          & 1.0901          & 1.2284          \\
ECAPA CNN-TDNN          & 0.9199          & 0.9253          & 0.9306          & 0.9465          & 0.9784          & 1.0422          & 1.1858          \\ \hline\hline
\textbf{MFA-TDNN (Standard)}       & \textbf{0.8615} & \textbf{0.8508} & \textbf{0.8402} & \textbf{0.8455} & \textbf{0.8615} & \textbf{0.9253} & \textbf{1.0210} \\ \hline
\end{tabular}
\vspace{-3.5mm}
\end{table*}

\vspace{-4mm}
\subsection{Training strategy}
The experiments are conducted using SpeechBrain Toolkit\footnote{https://speechbrain.github.io/}. For fair comparisons, all four systems discussed above are trained under the same training strategy following that in~\cite{desplanques2020ecapa}. Specifically, the Adam optimizer with cyclical learning rate~\cite{smith2017cyclical} varying between 1e-8 and 1e-3 following triangular policy~\cite{smith2017cyclical} is used for training all models. The loss function is additive angular margin softmax (AAM-softmax)~\cite{deng2019arcface} with a margin of 0.2 and a scale of 30. In addition, a weight decay used for all the weights in the model is set to 2e-5. 
 
All the samples are cut into 3-seconds segments during training. The mini-batch size is 384 (64 original samples each with 5 augmented samples). A total of 12 epochs are trained for each system. At the end of each epoch, the model is evaluated on the validation set to find the best model for testing.

\vspace{-2mm}
\subsection{Evaluation protocol}

We report the performances in terms of equal error rate (EER) and the minimum detection cost function (minDCF) with $P_{target}$ = 0.01 and $C_{FA}$ = $C_{Miss}$ = 1.
The test trial scores are produced by calculating the cosine distance between embeddings. 
The S-norm~\cite{kenny2010bayesian} is applied to normalize the scores.

\vspace{-3mm}
\section{Results and discussions}
\label{sec:results}

\tab~\ref{tab:sota} shows the performance comparison of the proposed MFA-TDNN based systems to the two baselines considered in this work. First, we observe that introduction of CNN module in baseline ECAPA-TDNN~\cite{desplanques2020ecapa} improves the performance in case of ECAPA CNN-TDNN~\cite{thienpondt2021integrating} baseline. On comparing to our proposed MFA-TDNN based systems, we observe that our MFA-TDNN (Standard) performs better than the stronger baseline ECAPA CNN-TDNN in most of the test cases. In addition, the lighter version of MFA-TDNN (Lite) also achieves improved performance than the ECAPA-TDNN baseline. 
It is noted that MFA-TDNN (Lite) has nearly 1.5M less parameters than MFA-TDNN (Standard). Further, it is worth to mention that both of our MFA-TDNN based proposed system has less parameters and multiply accumulate (MAC) operations than the two baselines. This suggests that our proposed systems not only provide improved results, but they also have compact architecture due to fewer parameters as well as computation complexity.         

\begin{figure}[]
\vspace{2mm}
\centerline{\includegraphics[scale=0.25]{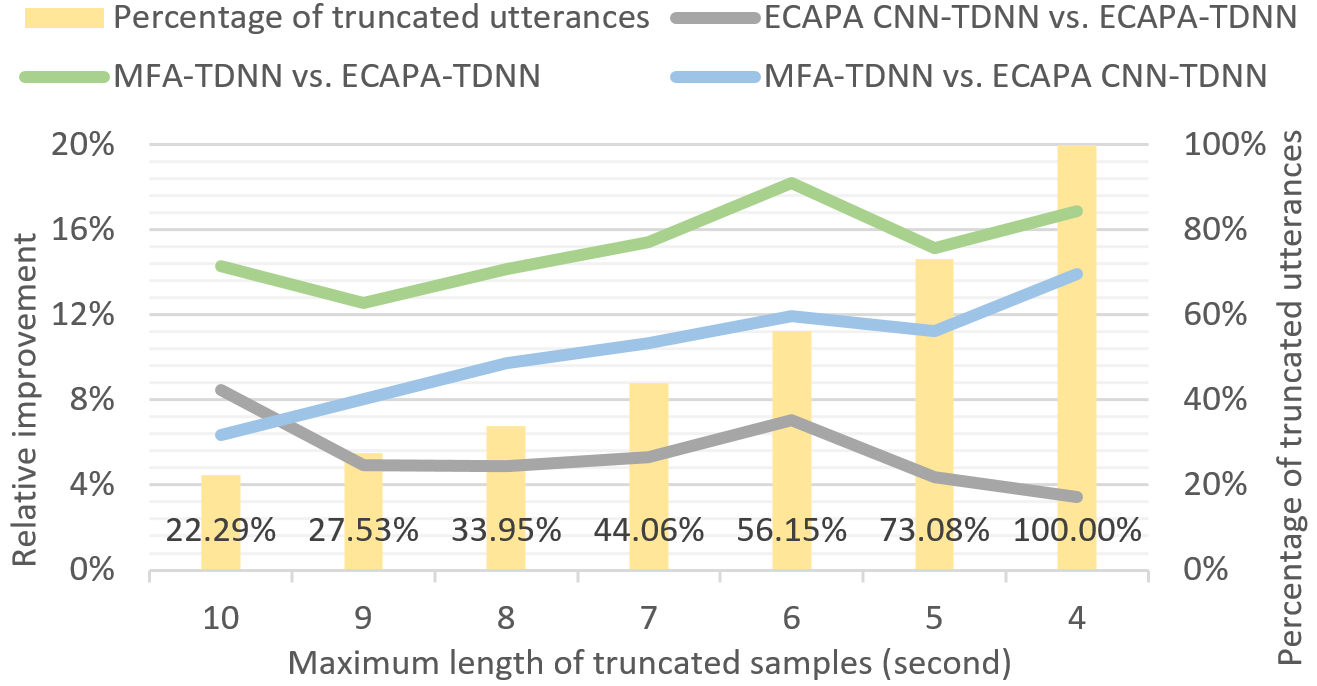}}
\caption{Relative improvement between different systems for short utterance test sets. The MFA-TDNN (Standard) is used.}
\label{fig:short}
 \vspace{-4mm}
\end{figure} 

Let us now compare the proposed systems with other existing SOTA systems. \tab~\ref{tab:sota} also shows the performances of several SOTA systems on the VoxCeleb database reported in respective works. 
We observe that MFA-TDNN outperforms all these SOTA systems. In addition, none of these SOTA systems is more efficient than MFA-TDNN systems when comparing the number of parameters and MACs. This further shows the advantages of MFA-TDNN systems due to the design of DM and FA modules.

We now compare our proposed MFA-TDNN with the two baseline systems under short utterance test scenario. The experiments are conducted on the truncated data sets as mentioned in Section~\ref{sec:Database} and the results are shown in \tab~\ref{tab:short}. All three systems are tested under different settings of maximum duration (4 to 10 seconds). From \tab~\ref{tab:short}, we observe that the MFA-TDNN consistently outperforms both the baselines ECAPA-TDNN and ECAPA CNN-TDNN for all the cases of short test utterances considered. It is worth noting that the improvement due to the proposed MFA-TDNN is particularly visible for short utterance scenarios.  

In order to better understand the performance trend of these systems under short utterance conditions, we make further analysis in Fig.~\ref{fig:short}. The curves in the figure represent the relative improvement trend for the three cases as the duration of samples shortens, while the bars indicate the percentage of testing utterances truncated. We observe that under the condition of maximum 10 seconds used for testing, 22.29\% testing samples are truncated. On the other hand, all the testing samples need to be shorten for 4-seconds test case. We also observe from Fig.~\ref{fig:short} that the relative improvement of ECAPA CNN-TDNN over ECAPA-TDNN (the grey curve) did not improve as the test duration continues to truncate. However, compared to ECAPA-TDNN and ECAPA CNN-TDNN, the relative improvements of the proposed MFA-TDNN is significantly visible as the duration of test utterances reduce. This shows the effectiveness of the proposed MFA-TDNN for text-independent SV with short utterances.

\vspace{-4mm}
\section{Conclusion}
\label{sec:Conclusion}
\vspace{-2mm}
We proposed a novel multi-scale frequency-channel attention (MFA) framework that efficiently captures the local information and frame-level temporal information by the dual-pathway multi-scale module and emphasize the important frequency region by frequency-channel attention in TDNN systems. Further, by cooperating these two novel modules, local cross-channel attention is achieved to model speaker characteristics from short samples efficiently. Studies on VoxCeleb dataset reveal that we greatly benefit from proposed MFA module. In addition, the further studies on the truncated test sets show the superiority of proposed MFA-TDNN for short sentences compared to various baselines.

\vfill\pagebreak


\small
\balance
\bibliographystyle{IEEEbib}

\bibliography{main}

\begin{thebibliography}{10}

\bibitem{campbell1997speaker}
Joseph~P Campbell,
\newblock ``Speaker recognition: A tutorial,''
\newblock {\em Proceedings of the IEEE}, vol. 85, no. 9, pp. 1437--1462, 1997.

\bibitem{tdref}
Matthieu H{\`e}bert,
\newblock ``Text-dependent speaker recognition,''
\newblock {\em Springer-Verlag Heidelberg}, pp. 743--762, 2008.

\bibitem{liu2022neural}
Tianchi Liu, Rohan~Kumar Das, Kong~Aik Lee, and Haizhou Li,
\newblock ``Neural acoustic-phonetic approach for speaker verification with
  phonetic attention mask,''
\newblock {\em IEEE Signal Processing Letters}, 2022.

\bibitem{das2018speaker}
Rohan~Kumar Das and S.~R.~Mahadeva Prasanna,
\newblock ``Speaker verification from short utterance perspective: a review,''
\newblock {\em IETE Technical Review}, vol. 35, no. 6, pp. 599--617, 2018.

\bibitem{rkd_thesis}
Rohan~Kumar Das,
\newblock {\em Speaker verification using sufficient train and limited test
  data},
\newblock PhD thesis, Septemebr 2017.

\bibitem{liu2020text}
Kai Liu and Huan Zhou,
\newblock ``Text-independent speaker verification with adversarial learning on
  short utterances,''
\newblock in {\em Proc. IEEE ICASSP}, 2020, pp. 6569--6573.

\bibitem{bhattacharya2017deep}
Gautam Bhattacharya, Md~Jahangir Alam, and Patrick Kenny,
\newblock ``Deep speaker embeddings for short-duration speaker verification.,''
\newblock in {\em Proc. Interspeech}, 2017, pp. 1517--1521.

\bibitem{huang2018angular}
Zili Huang, Shuai Wang, and Kai Yu,
\newblock ``Angular softmax for short-duration text-independent speaker
  verification.,''
\newblock in {\em Proc. Interspeech}, 2018, pp. 3623--3627.

\bibitem{snyder2017deep}
David Snyder, Daniel Garcia-Romero, Daniel Povey, and Sanjeev Khudanpur,
\newblock ``Deep neural network embeddings for text-independent speaker
  verification.,''
\newblock in {\em Proc. Interspeech}, 2017, pp. 999--1003.

\bibitem{lee2021xi}
Kong~Aik Lee, Qiongqiong Wang, and Takafumi Koshinaka,
\newblock ``Xi-vector embedding for speaker recognition,''
\newblock {\em IEEE Signal Processing Letters}, pp. 1385--1389, 2021.

\bibitem{liu2019unified}
Tianchi Liu, Maulik~C Madhavi, Rohan~Kumar Das, and Haizhou Li,
\newblock ``A unified framework for speaker and utterance verification.,''
\newblock in {\em Proc. Interspeech}, 2019, pp. 4320--4324.

\bibitem{yi2021perceptual}
Yi~Ma, Kong~Aik Lee, Ville Hautam{\"a}ki, and Haizhou Li,
\newblock ``{PL-EESR}: Perceptual loss based end-to-end robust speaker
  representation extraction,''
\newblock in {\em Proc. IEEE ASRU}, 2021.

\bibitem{zhou2021resnext}
Tianyan Zhou, Yong Zhao, and Jian Wu,
\newblock ``{ResNeXt and Res2Net} structures for speaker verification,''
\newblock in {\em IEEE Spoken Language Technology Workshop (SLT)}, 2021, pp.
  301--307.

\bibitem{li2017deep}
Chao Li, Xiaokong Ma, Bing Jiang, Xiangang Li, Xuewei Zhang, Xiao Liu, Ying
  Cao, Ajay Kannan, and Zhenyao Zhu,
\newblock ``Deep speaker: an end-to-end neural speaker embedding system,''
\newblock {\em arXiv preprint arXiv:1705.02304}, 2017.

\bibitem{desplanques2020ecapa}
Brecht Desplanques, Jenthe Thienpondt, and Kris Demuynck,
\newblock ``{ECAPA-TDNN}: Emphasized channel attention, propagation and
  aggregation in {TDNN} based speaker verification,''
\newblock in {\em Proc. Interspeech}, 2020, pp. 3830--3834.

\bibitem{zhou2020dynamic}
Dao Zhou, Longbiao Wang, Kong~Aik Lee, Yibo Wu, Meng Liu, Jianwu Dang, and
  Jianguo Wei,
\newblock ``Dynamic margin softmax loss for speaker verification.,''
\newblock in {\em Proc. Interspeech}, 2020, pp. 3800--3804.

\bibitem{chung2020defence}
Joon~Son Chung, Jaesung Huh, Seongkyu Mun, Minjae Lee, Hee~Soo Heo, Soyeon
  Choe, Chiheon Ham, Sunghwan Jung, Bong-Jin Lee, and Icksang Han,
\newblock ``In defence of metric learning for speaker recognition,''
\newblock in {\em Proc. Interspeech}, 2020, pp. 2977--2981.

\bibitem{tao2020audio}
Ruijie Tao, Rohan~Kumar Das, and Haizhou Li,
\newblock ``Audio-visual speaker recognition with a cross-modal discriminative
  network,''
\newblock in {\em Proc. Interspeech}, 2020, pp. 2242--2246.

\bibitem{thienpondt2021integrating}
Jenthe Thienpondt, Brecht Desplanques, and Kris Demuynck,
\newblock ``Integrating frequency translational invariance in {TDNNs} and
  frequency positional information in {2D ResNets} to enhance speaker
  verification,''
\newblock {\em arXiv preprint arXiv:2104.02370}, 2021.

\bibitem{gao2019res2net}
Shanghua Gao, Ming-Ming Cheng, Kai Zhao, Xin-Yu Zhang, Ming-Hsuan Yang, and
  Philip~HS Torr,
\newblock ``Res2net: A new multi-scale backbone architecture,''
\newblock {\em IEEE transactions on pattern analysis and machine intelligence},
  pp. 652--662, 2019.

\bibitem{liu2020speaker}
Tianchi Liu, Rohan~Kumar Das, Maulik Madhavi, Shengmei Shen, and Haizhou Li,
\newblock ``Speaker-utterance dual attention for speaker and utterance
  verification,''
\newblock in {\em Proc. Interspeech}, 2020, pp. 4293--4297.

\bibitem{jiang2019effective}
Yiheng Jiang, Yan Song, Ian McLoughlin, Zhifu Gao, and Li-Rong Dai,
\newblock ``An effective deep embedding learning architecture for speaker
  verification,''
\newblock in {\em Proc. Interspeech}, 2019, pp. 4040--4044.

\bibitem{hu2018squeeze}
Jie Hu, Li~Shen, and Gang Sun,
\newblock ``Squeeze-and-excitation networks,''
\newblock in {\em Proc. IEEE CVPR}, 2018, pp. 7132--7141.

\bibitem{wang2020eca}
Qilong Wang, Banggu Wu, Pengfei Zhu, Peihua Li, Wangmeng Zuo, and Qinghua Hu,
\newblock ``{ECA-Net}: Efficient channel attention for deep convolutional
  neural networks,''
\newblock in {\em Proc. IEEE CVPR}, 2020, pp. 11531--11539.

\bibitem{nagrani2017voxceleb}
Arsha Nagrani, Joon~Son Chung, and Andrew Zisserman,
\newblock ``{VoxCeleb}: a large-scale speaker identification dataset,''
\newblock in {\em Proc. Interspeech}, 2017, pp. 2616--2620.

\bibitem{chung2018voxceleb2}
Joon~Son Chung, Arsha Nagrani, and Andrew Zisserman,
\newblock ``{VoxCeleb2}: Deep speaker recognition,''
\newblock in {\em Proc. Interspeech}, 2018, pp. 1086--1090.

\bibitem{park2019specaugment}
Daniel~S Park, William Chan, Yu~Zhang, Chung-Cheng Chiu, Barret Zoph, Ekin~D
  Cubuk, and Quoc~V Le,
\newblock ``Specaugment: A simple data augmentation method for automatic speech
  recognition,''
\newblock in {\em Proc. Interspeech}, 2019, pp. 2613--2617.

\bibitem{ko2015audio}
Tom Ko, Vijayaditya Peddinti, Daniel Povey, and Sanjeev Khudanpur,
\newblock ``Audio augmentation for speech recognition,''
\newblock in {\em Proc. Interspeech}, 2015, pp. 3586--3589.

\bibitem{zhang2020aret}
Ruiteng Zhang, Jianguo Wei, Wenhuan Lu, Longbiao Wang, Meng Liu, Lin Zhang,
  Jiayu Jin, and Junhai Xu,
\newblock ``{ARET:} aggregated residual extended time-delay neural networks for
  speaker verification,''
\newblock in {\em Proc. Interspeech}, 2020, pp. 946--950.

\bibitem{kwon2021ins}
Yoohwan Kwon, Hee-Soo Heo, Bong-Jin Lee, and Joon~Son Chung,
\newblock ``The ins and outs of speaker recognition: lessons from voxsrc
  2020,''
\newblock in {\em Proc. IEEE ICASSP}, 2021, pp. 5809--5813.

\bibitem{smith2017cyclical}
Leslie~N Smith,
\newblock ``Cyclical learning rates for training neural networks,''
\newblock in {\em IEEE winter conference on applications of computer vision
  (WACV)}, 2017, pp. 464--472.

\bibitem{deng2019arcface}
Jiankang Deng, Jia Guo, Niannan Xue, and Stefanos Zafeiriou,
\newblock ``Arcface: Additive angular margin loss for deep face recognition,''
\newblock in {\em Proc. IEEE CVPR}, 2019, pp. 4690--4699.

\bibitem{kenny2010bayesian}
Patrick Kenny,
\newblock ``Bayesian speaker verification with heavy-tailed priors.,''
\newblock in {\em Proc. Odyssey}, 2010, p.~14.

\end{thebibliography}
\end{document}